\documentclass[12pt,preprint]{aastex}
\def\msun{{\rm ~M}_{\odot}}

\usepackage{graphicx}

\begin{document}

\title{On the Transition from Accretion Powered to Rotation Powered Millisecond Pulsars}
\author{J. Takata\altaffilmark{1}
\email{takata@hku.hk}
K. S. Cheng\altaffilmark{1}
\email{hrspksc@hkucc.hku.hk}
\and
Ronald E. Taam\altaffilmark{2,3}
\email{r-taam@northwestern.edu}}
\altaffiltext{1}{Department of Physics, University of Hong Kong, Pokfulam Road, Hong Kong}
\altaffiltext{2}{Department of Physics and Astronomy, Northwestern University,
2131 Tech Drive, Evanston, IL 60208}
\altaffiltext{3}{Academia Sinica Institute of Astronomy and Astrophysics - TIARA,
P.O. Box 23-141, Taipei, 10617 Taiwan}

\begin{abstract}
The heating associated with the deposition of $\gamma$-rays in an accretion disk is proposed 
as a mechanism to facilitate the transformation of a low mass X-ray binary to the radio millisecond 
pulsar phase.  The $\gamma$-ray emission produced in the outer gap accelerator in the pulsar magnetosphere 
likely irradiates the surrounding disk, resulting in its heating and to the possible escape of matter 
from the system. We apply the model to PSR J1023+0038, which has recently been discovered as a newly 
born rotation powered millisecond pulsar. The predicted $\gamma$-ray luminosity $\sim 6 \times 
10^{34}~\mathrm{erg~s^{-1}}$ can be sufficient to explain the disappearance of the truncated disk existing 
during the 8~month$\sim 2$~yr period prior to the 2002 observations of J1023+0038 and the energy input 
required for the anomalously bright optical emission of its companion star.
\end{abstract}

\section{Introduction}

The discovery of pulsed radio emission at 1.69 ms from the source J102347.67+003841.2 (see 
Archibald et al. 2009), offers a unique opportunity to study the evolution of a system from the 
low mass X-ray binary (LMXB) phase to the recycled millisecond pulsar (MSP) phase.  The system 
has an orbital period of 4.75 hrs (see Woudt, Warner, \& Pretorius 2004) and is distinguished 
from other systems by a change in optical appearance from a spectrum characterized by broad double 
peaked hydrogen and helium emission lines (see Bond et al. 2002) to a late type absorption 
spectrum (see Thorstensen \& Armstrong 2005).  Further analysis based on SDSS spectra by Wang 
et al. (2009) provide evidence for the presence of a truncated accretion disk in the early 
observations in 2001 and its possible absence in 2002. The recent detection of X-ray pulsations 
by Archibald et al. (2010) suggests that pulsar high-energy emission activity was already present in 2008.

The description of the transition from the accretion powered to the rotation powered phase is 
not understood due to the complexities in the description of the interaction between the 
magnetosphere of a neutron star (NS) and its accretion disk. Simple arguments have been discussed 
in early work by Campana et al. (1998).  Although theoretical developments have not 
progressed significantly, recent observations of the optical counterparts of 
NS soft X-ray transients in the quiescent state have provided indirect evidence 
for additional heating of the companion star.  As examples, orbital modulations in the 
optical emission from J102347.68+003841.2 (Thorstensen \& Armstrong 2005), SAX J1808.4-3658 
(Burderi et al. 2003; Deloye et al. 2008), XTE 1814-338 (D'Avanzo et al. 2009) and IGR J00291+5934 
(Jonker, Torres \& Steeghs 2008) have been found in their respective quiescent states. Specifically, 
the amplitude of the modulation cannot be explained by irradiation associated with the X-ray emission 
from the disk or NS due to insufficient luminosity, indicating the need for the operation 
of an additional heating source. To provide an explanation of the orbital modulation of the optical 
emission, pulsar wind models have been suggested in which heating of the donor is due to the 
effect of a relativistic pulsar wind (Burderi et al. 2003).  In addition, the interaction between 
the pulsar and its companion has been suggested as leading to the formation of an intrabinary 
shock (see Campana et al. 1998) for the generation of non-thermal X-ray radiation observed in 
J0024-7204W in the globular cluster 47 Tuc (see Bogdanov, Grindlay \& Berg 2005), in PSR J1740-5340 
in NGC 6397 (Grindlay, et al. 2002) and in PSR~J1023+0038 (Archibald et al. 2010).  Finally, the 
disappearance of the truncated disk in J102347.67+003841.2 may have been a direct result of the 
activation of the pulsar. 
  
In order to facilitate this transition, the accretion of matter onto the NS  must 
decrease rapidly, for otherwise, the NS will spin down (e.g., see Jeffrey 1986).  
In contrast to the case of radio MSPs in long orbital period systems where a red 
giant companion can detach from its Roche lobe as its envelope is exhausted, a sudden decrease in 
the mass transfer from the donor is not generally expected for the donors in short period binary 
systems.  For these systems, the disk must be truncated sufficiently such that the mass flow is 
prevented from entering into the NS magnetosphere.  A mechanism possibly leading to a 
sudden decrease in the accretion flow is the operation of the so called "propeller" effect (Campana 
et al. 1998; Romanova et al.  2009), in which the Alfven radius exceeds the corotation radius, during 
the quiescent state of a X-ray transient LMXB phase.  In this regime, however, the prevention of 
matter accreting onto the NS is uncertain. That is, matter can either be swept away 
or accrete intermittently onto the NS.  Ablation due to the 
action of a pulsar wind may also facilitate the removal of the disk around the pulsar (Wang et al. 
2009), however, an important theoretical uncertainty is the description by which a 
particle kinetic dominated flow can be formed near the pulsar (Kirk \& Skj\ae raasen 2003; Arons 2008).

In this Letter, we explore the possibility that irradiation by $\gamma$-ray emission produced 
in the outer gap accelerator within a pulsar magnetosphere can facilitate dissolution of the 
disk. In \S 2, we briefly describe the theoretical concepts of the model and apply the theory 
to J102347.67+003841.2 in \S 3. In the last section, we discuss the implications of our model 
and conclude.

\section{Theoretical Model}

\begin{figure}[t]
\epsscale{.40}
\plotone{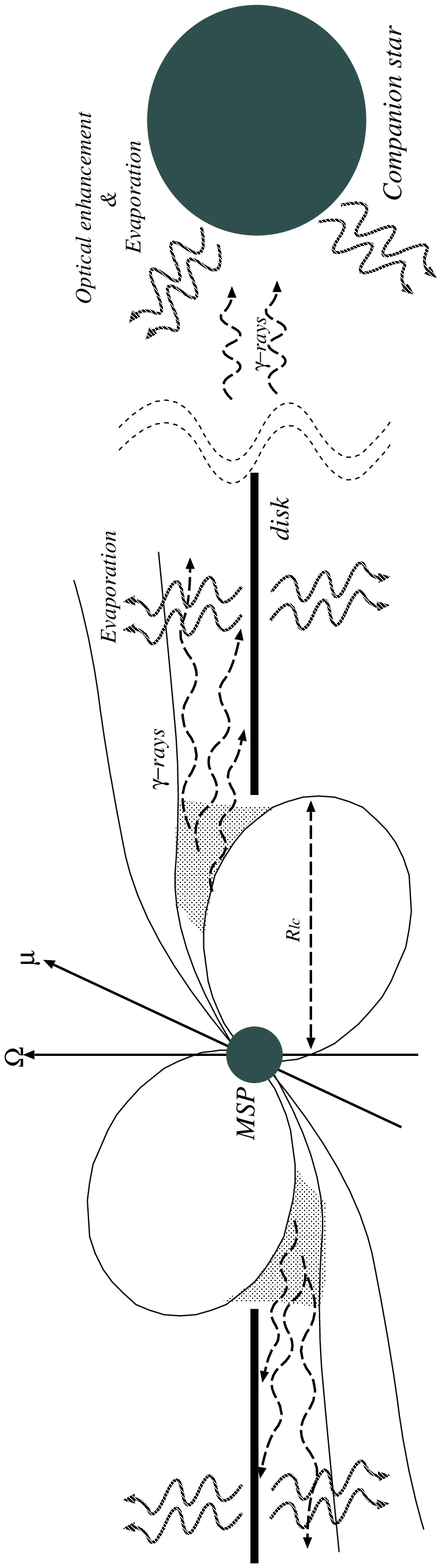}
\caption{The schematic view of irradiation by the 
$\gamma$-rays from the outer gap 
(shadowed region) within a  pulsar magnetosphere. The outer gap accelerator 
is extending between the null charge surface, where 
$\vec{\Omega}\cdot \vec{B}=0$, to the light cylinder.}
\label{dipole}
\end{figure}

The activation of the radio pulsar phase of a NS in a short period LMXB is likely to 
occur during the quiescent state (see Campana et al. 1998) when the mass accretion rates are 
lower than the time averaged value, typically $\dot{M}\sim 10^{14-16}~\mathrm{g/s}$ (Heinke 
et al. 2009).  In such a phase, the magnetosphere can extend beyond the light cylinder radius. 
If the pulsar magnetosphere is sufficiently clear of matter, $\gamma$-ray photons with an energy 
$\sim$ GeV can be  produced in an outer gap accelerator (Cheng, Ho \& Ruderman 1986a, 1986b) 
via the curvature radiation process of high-energy particles.
The emitted $\gamma$-rays can irradiate the accretion disk to promote the escape of disk material 
from the system (c.f. Figure~\ref{dipole}). Sufficient heating of the disk may directly lead to 
the escape of matter or to an increase in its vertical scale height.  In this case, the disk would 
subtend a larger solid angle as seen from the NS, thereby facilitating the acceleration of disk 
material associated with the radiation pressure of the magnetic dipole radiation.  In the following, 
we describe the process. 

\subsection{$\gamma$-ray emissions from the outer gap}

In the outer gap model for the rotation powered pulsars, particles can be accelerated by the 
electric field along the magnetic field lines in the region where the local charge density 
deviates from Goldreich-Julian charge density.  The typical strength of the accelerating field 
in the gap is expressed as (Zhang \& Cheng 1997)
\begin{equation}
E_{||}\sim \frac{f^2V_{a}}{s}\sim 9.2\times 10^5f^2 \left(\frac{P}{10^{-3}~\mathrm{s}}\right)^{-3}
\left(\frac{B_s}{10^{8}~\mathrm{G}}\right)\left(\frac{s}{R_{lc}}\right)^{-1}
\left(\frac{R_s}{10^{6}~\mathrm{cm}}\right)^3 \mathrm{(c.g.s.)},
\label{electric}
\end{equation}
where $V_a=B_sR_s^3/R_{lc}^2$ is the electrical potential drop on the polar cap, $P$ is 
the rotation period, $B_s$ is the stellar magnetic field, $s$ is the curvature radius of the 
magnetic field line, $R_{lc}=cP/2\pi$ is the light cylinder radius, and $R_s$ is the stellar radius. 
In addition, $f$ is the fractional gap thickness, which is defined by the ratio of the gap 
thickness to $R_{lc}$.  By assuming force balance between the 
acceleration and radiation reaction of the curvature radiation process, the typical Lorentz 
factor of the particles in the gap is
\begin{equation}
\Gamma=\left(\frac{3s^2}{2e}E_{||}\right)^{1/4}\sim 1.6\times 10^7 f^{1/2}P^{-1/4}_{-3}B_8^{1/4}s_{1}^{1/4}R_{6}^{3/4},
\end{equation}
where $P_{-3}=(P/1~\mathrm{ms})$,  $B_8=(B_s/10^8~\mathrm{G})$, $s_1=(s/R_{lc})$, and $R_6=(R_s/10^6~
\mathrm{cm})$.  The corresponding photon energy of the curvature radiation is
\begin{equation}
 E_{\gamma}=\frac{3hc\Gamma^3}{4\pi s}\sim 25f^{3/2}P_{-3}^{-7/4}B_{8}^{3/4}s_{1}^{-1/4}R_6^{9/4}~\mathrm{GeV}.
\label{energy}
\end{equation}
The $\gamma$-rays emitted in the gap will be converted into electron/positron pairs via the 
photon-photon pair-creation process with X-ray photons from either the NS surface or the 
accretion disk.  Because the created pairs 
tend to screen the electric field in the gap, the gap thickness is controlled by the photon-photon 
pair-creation process. Using the pair-creation condition that $E_{\gamma}E_{X}\sim (m_ec^2)^2$, 
$f$ is estimated as 
\begin{equation}
f\sim 0.22P^{7/6}_{-3}B^{-1/2}_{8}s_1^{1/6}R_6^{-3/2}E_{0.1}^{-2/3},
\end{equation}
where $E_{0.1}$ is the typical X-ray photon energy in units of 100~eV. The luminosity of the 
$\gamma$-ray emission from the outer gap is estimated as
\begin{equation}
L_{\gamma}\sim f^3 L_{sd}\sim 4.2\times 10^{33} P^{-1/2}_{-3}B^{1/2}_{8}
s_1^{1/2}R_6^{3/2}E_{0.1}^{-2}~\mathrm{erg/s}, 
\label{glumi}
\end{equation}
where $L_{sd}=4(2\pi)^4B^2R^6/6c^3P^4$ is the pulsar spin down power.

The rotation period of the newly switched on MSP is related to the history of the 
accretion onto the NS.  In particular, its rotation period in the accretion stage may 
be related to the equilibrium spin period, which is obtained by equating the co-rotation radius,
$r_{co}=(GMP^2/4\pi^2)^{1/3}$ to the magnetospheric radius, $r_M\sim 2.4 \times 10^6 
B_8^{4/7}R_6^{12/7}\dot{M}_{15}^{-2/7}M_{1.4}^{-1/7}$~cm.  Here, $M$ is the NS mass, 
$\dot{M}_{15}$ is the time averaged accretion rate in units of $10^{15}$~g/s and $M_{1.4}$ is 
the NS  mass in units $1.4 \msun$. However, the spin period may not be equal to 
its equilibrium spin period since the pressure build up associated with the accumulation of 
disk matter at the magnetospheric boundary can push the magnetospheric boundary inside the 
magnetospheric radius $r_M$ (Romanova et al. 2009), resulting in a higher spin period than its 
equilibrium value.  Because of this uncertainty, we introduce a parameter, $\beta$, as $r_{co}=
\beta r_{M}$, which yields the rotation period,  
\begin{equation}
P_{e,-3}=1.7\kappa B_8^{6/7}.
\label{period}
\end{equation}
where  $\kappa=\beta^{3/2}R_6^{18/7}\dot{M}_{15}^{-3/7}M_{1.4}^{-5/7}$. The spin period is 
illustrated in Fig.~\ref{pb} as a function of the stellar magnetic field with $\kappa=1$ (solid line), 
$0.4$ (dashed line) and $0.2$ (dashed-dotted line). For example, with $\beta=0.5$ and $R_6=M_{1.4}=1$, 
$\kappa=1,~0.4$ and 0.2 corresponding to $\dot{M_{15}}\sim 0.1,\sim 0.7$ and $\sim 4$.
In addition, we also plot the observed spin periods and inferred magnetic fields of the radio 
MSP population in the Galactic field. Only MSPs in binaries with orbital 
periods less than about 10 days are included since those systems with longer periods may have had 
red giant companions. In this case, the dissolution of the disk is due to the cessation of mass 
transfer as the envelope of the giant star is exhausted leading to the shrinkage of the red giant 
within its Roche lobe.
 
\subsection{$\gamma$-ray irradiation of the disk}
The magnetospheric $\gamma$-rays irradiating the surrounding disk may be absorbed via the so called 
pair-creation process in a Coulomb field by nuclei in the accretion disk. Further absorption 
will occur as the relativistic pairs created with a Lorentz factor $\Gamma\sim 10^{3}$ will 
transfer their energy and momentum to the disk matter via the Coulomb scattering process.   

The cross section of the above pair-creation process is given by Lang (1999) as
\begin{equation}
\sigma(E_{\gamma})=3.5\times 10^{-3}Z^2\sigma_T
\left[\frac{7}{9}\log \left(\frac{183}{Z^{1/2}}\right)-\frac{1}{53}\right]
~~\mathrm{for}~~\frac{E_{\gamma}}{2}>>\frac{m_ec^2}{\alpha Z^{1/3}}, 
\end{equation}
where $\sigma_T$ is the Thomson cross section,  $Z$ is the atomic number, $\alpha$ is the 
fine structure constant, and $m_ec^2$ is electron rest mass energy. All $\gamma$-rays 
irradiating the disk can be absorbed if the column density of the accretion disk exceeds 
a critical value given by
\begin{equation}
\Sigma_{crit}\sim \frac{m_p}{\sigma} \frac{h}{l} \sim 60\frac{h}{l}~\mathrm{g/cm^2},
\label{scrit}
\end{equation}
where $l$ is the propagation length of the $\gamma$-rays in the disk, $h$ is the thickness of 
the disk, $Z^2=3$ appropriate for a solar abundance and $m_p$ is the mass of a proton. 
It can be seen that the above column density is smaller than the typical column density of 
the accretion disk.  For the standard gas pressure supported 
disk model (Frank, King \& Raine, 2002), the column density is
\begin{equation}
\Sigma(R_{lc})\sim 2.1\times 10^{3}\alpha_{0.1}^{-4/5}\dot{M}_{15}^{7/10}
M_{1.4}^{1/4}P^{-3/4}_{-3}~\mathrm{g/cm^2}, 
\end{equation}
where $\alpha_{0.1}$ is the viscosity parameter in units of 0.1. We find that the required column  density~(\ref{scrit}) is much smaller than that of the standard disk model for viscosity parameters 
less than unity. 

An estimate of a critical accretion rate below which the radiation associated with $\gamma$-rays 
from the outer gap can energetically eject the matter can be obtained by equating $L_{\gamma}=
GM\dot{M}/2R_{lc}$ with equation~(\ref{glumi}), given by 
\begin{equation}
\dot{M}_{c}=2.1\times 10^{14}P_{-3}^{1/2}B_8^{1/2}s_1^{1/2}R_6^{3/2}E_{0.1}^{-2}
M_{1.4}^{-1}~\mathrm{g/s}.
\label{mcrit}
\end{equation}
  
\begin{figure}[t]
\epsscale{.80}
\plotone{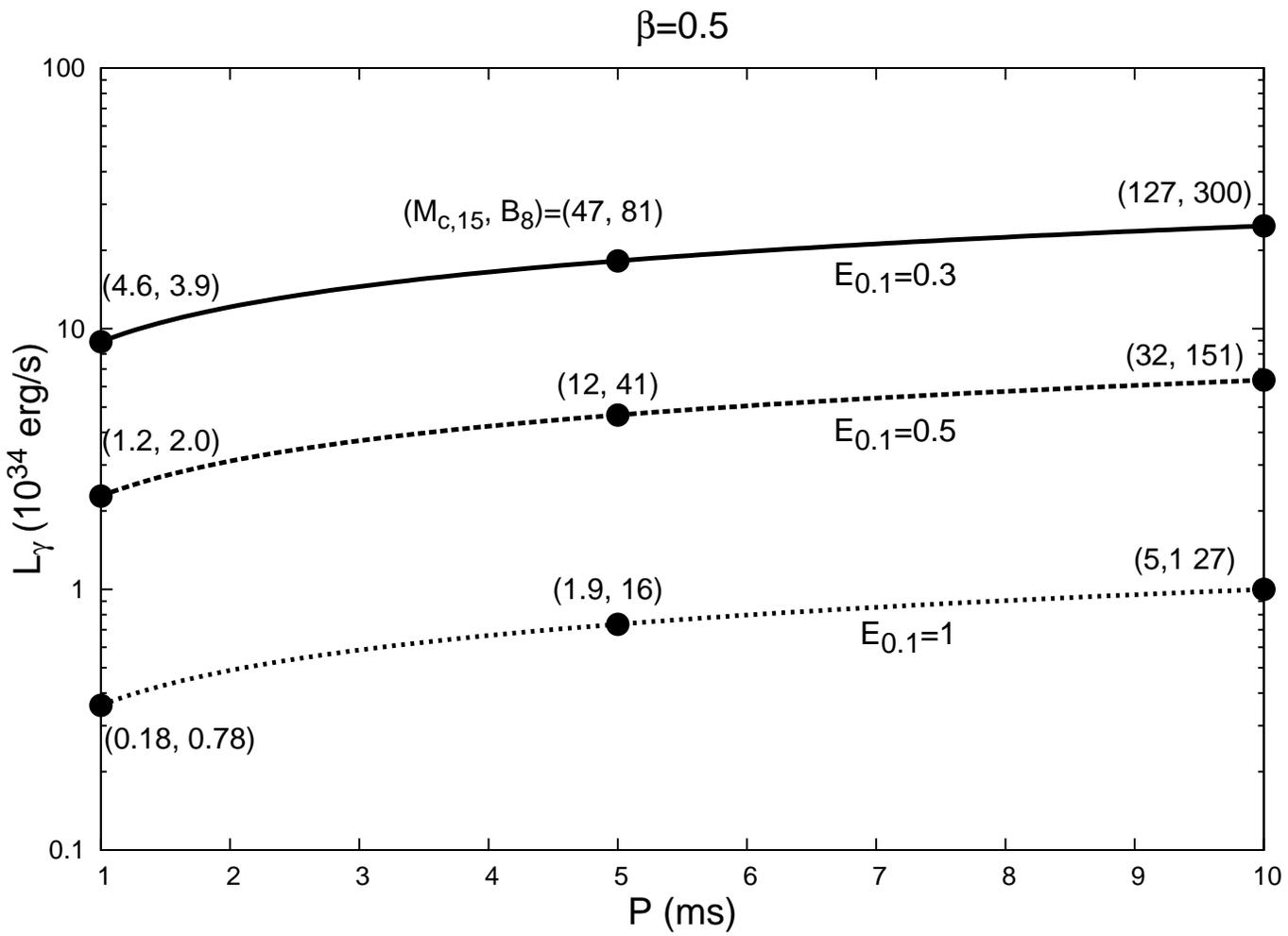}
\caption{The predicted $\gamma$-ray luminosity and the surface magnetic field at the transition 
from the accretion powered to rotation powered phases. The results are for the accretion rate described 
by the critical value, that is,  $\dot{M}=\dot{M}_{c}$ and $\beta=0.5$.  The different lines represent 
results for the different typical energy of the X-ray photons.} 
\label{lp}
\end{figure}

Applying the critical accretion rate~(\ref{mcrit}) and $\beta=0.5$ to equations~(\ref{period}) and
using equation~(\ref{glumi}), the predicted $\gamma$-ray luminosity and the surface magnetic field 
are illustrated  as a function of equilibrium period and the X-ray photon energy in Fig.~\ref{lp}.  
The different lines represent the results for the different X-ray photon energies.  For a pulsar 
with $P_{-3}=5$ and $E_{0.1}=0.5$, for example, the predicted $\gamma$-ray luminosity and the surface 
magnetic field are $L_{\gamma}\sim 5\times 10^{34}~\mathrm{erg~s^{-1}}$ and $B_8=41$, respectively.
Note that if the true accretion rate is less than the critical accretion rate, the predicted magnetic 
field and $\gamma$-ray luminosity for the specific set of the rotation period and the  X-ray photon 
energy provide smaller value than those associated with the critical rate.

\section{Application to PSR J1023+0038}

\begin{figure}[t]
\epsscale{.80}
\plotone{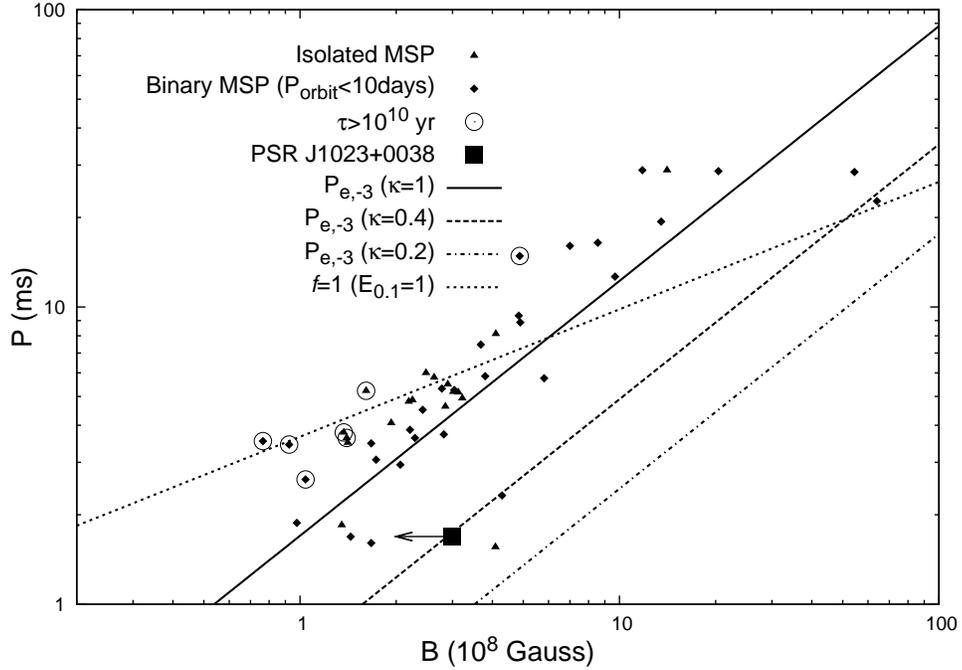}
\caption{Plot of $P$ vs. $B_s$ for the MSPs. The triangles and diamonds  represent the field 
isolated MSPs and the field binary MSP with an orbital period smaller than 10~days. In addition, 
the circles shows the MSPs with a characteristic age larger than 10~Gyr. The filled square 
represents the newly born MSP J1023+0038 with $B_s=3 \times 10^8$~Gauss, which is indicated as an 
upper limit. The solid, dashed and dotted-dashed line present the relation $P_e=1.7\kappa B^{7/6}$ with 
$\kappa=1$, $\kappa=0.4$ and $\kappa=0.2$. The dotted line represents the line of $f=1$ with 
$E_{0.1}=1$. The results are for $R_6=M_{1.4}=s_1=1$.}
\label{pb}
\end{figure}

As an application of our model, we estimate the $\gamma$-ray luminosity which inhibits accretion 
into the magnetosphere, once the rotation powered pulsar is activated.  In Fig.~\ref{pb}, the observed 
characteristics (filled box) for PSR J1023+0038 are indicated.  For the X-ray emission, Homer et al. 
(2006) found that the X-ray emission in 2004 observations are well fit by a NS atmosphere model 
($E_{X}\sim 30$~eV) plus power law component. However,  we choose the X-ray emission from the NS 
stellar surface rather than the X-ray emission from the pair-creation originating from either the 
pulsar magnetosphere or a shock where material overflowing the companion interacts with the pulsar wind. 
If the X-ray emission is magnetospheric in origin, the synchrotron emission from the secondary pairs 
produced outside  the main acceleration region is possible. In such a case, because the high-energy
 photons are propagating to convex side of the magnetic field lines, the non-thermal X-ray photons are 
unlikely to illuminate the gap region and, therefore, collisions with the $\gamma$-rays inside 
the gap are not expected.  On the other hand, if the non-thermal X-rays originate from the shock 
region, the photon-number density in the pulsar magnetosphere is much smaller than that associated 
with the thermal X-rays from the NS surface, indicating the photon-photon pair-creation 
process in the gap mainly occurs with the thermal X-rays. 

As Fig.~\ref{pb} reveals, the parameter $\kappa\le 0.4$ in the relation~(\ref{period}) can be consistent 
with the observed rotation period and upper limit of the inferred surface magnetic field.  This 
may indicate that the averaged accretion rate is order of $\dot{M}\sim  10^{15}$~g/s unless $\beta$ 
is much smaller than unity.  On the other hand, the predicted $\gamma$-ray luminosity~(\ref{glumi}) and the critical 
accretion rate~(\ref{mcrit}) give  $L_{\gamma}\sim 6\times 10^{34}~\mathrm{erg~s^{-1}}$ and $\dot{M}\sim 
5\times 10^{15}~\mathrm{g/s}$, respectively, with  $B_8=3$ and $E_{0.1}=0.3$ and $s_1=R_6=M_{1.4}=1$, 
indicating the $\gamma$-ray irradiation on the disk would be sufficient to inhibit the accretion flow 
into the magnetosphere.  We note that the predicted luminosity $L_{\gamma}\sim 6\times 10^{34}~
\mathrm{erg~s^{-1}}$ is consistent with the required luminosity from an irradiating source to explain 
the heating of the companion star (Thorstensen \& Armstrong 2005). 

As argued by Wang et al. (2009), there is evidence that a truncated disk existed in SDSS 
J102347.6+00381 during the 8~month$\sim$ 2 year period prior to the 2002 observations.  
Based on a standard disk model a disk extending from  $10^9$~cm to $5.7\times 10^{10}$~cm 
with a total disk mass of $M_d\sim 1.7\times 10^{23}$~g was inferred.  Here, we examine 
whether the purported $\gamma$-ray irradiation can remove the disk within a time scale of 
less than 2 yr.  

Wang et al. (2009) estimate a disk column density of $\Sigma_d\sim 
36(\alpha/0.1)^{-4/5}\dot{M}_{16}^{7/10} r_{10}^{-3/4}$, which is higher than the required 
density~(\ref{scrit}) for $\gamma$-ray absorption because the propagation distance in the disk is much 
longer than the disk thickness. Here, we examine the fraction 
of the $\gamma$-ray power which can irradiate the disk at $r=10^{9}\sim 10^{10}$~cm.  If we apply 
(1) a pure dipole magnetic field, (2) the geometry of the outer gap accelerator, which extends 
from the so called Goldreich-Julian null charge surface, 
where the product $\vec{\Omega}\cdot \vec{B}=0$, to the light cylinder, and (3) a 
standard disk model, implying a disk thickness of $h\sim 10^{8}$~cm at $r\sim 10^{10}$~cm 
from the pulsar, we find that about $\eta\sim 0.1~\%$ of the outer gap region can illuminate 
the disk extending beyond $r\ge 10^{9}$~cm. Assuming the $\gamma$-ray irradiation can remove 
the matter at a rate of $\dot{M}=\eta L_{\gamma}r/GM$, the time scale of the life of the 
disk, $\tau\sim M_d/\dot{M}$, is 
\begin{equation}
\tau\sim 1\left(\frac{\eta L_{\gamma}}{6\cdot 10^{31}~\mathrm{erg/s}}\right)^{-1}
\left(\frac{r}{10^{10}~\mathrm{cm}}\right)^{-1}
\left(\frac{M_d}{10^{23}~\mathrm{g}}\right)M_{1.4}~\mathrm{yr},
\end{equation}
which can be consistent with the life time inferred from the observation. 

\section{Discussion \& Conclusion}

Since our scenario for describing the phase from the accreting MSP phase to the radio MSP phase 
invokes the $\gamma$-ray radiation from the outer gap, observations using the $Fermi$ satellite 
provide important constraints on the $\gamma$-ray emission model.  Since the spin down power of 
PSR J1023+0038 ($L_{sd} \le 3 \times 10^{35}$~erg/s) is relatively high compared to $\gamma$-ray 
emitting pulsars detected by $Fermi$ (Abdo et al. 2009) and its distance ($d\sim 1$~kpc) is similar 
to these $Fermi$ sources, PSR J1023+0038 is a prime candidate for $\gamma$-ray emission. 

Recently Tam et al. (2010) report the discovery of the $\gamma$-ray emissions in the 
direction of PSR J1023+0038 with a flux $5\times 10^{-12}~\mathrm{erg~cm^{-2} s^{-1}}$. Although 
the pulsation has not been confirmed in $\gamma$-rays by Tam et al. (2010), its position is 
in coincidence with the radio position, which lies  within the 68\% confidence-level error circle 
of the $\gamma$-ray position.  The $\gamma$-ray flux can be estimated as $F_{\gamma}\sim 
L_{\gamma}/\delta\Omega d^2\sim 10^{-9}~\mathrm{erg/cm^2 s}$, where $L_{\gamma}=6\times 
10^{34}~\mathrm{erg~s^{-1}}$, the typical solid angle $\delta\Omega=2$, and the distance $d=1.3$~kpc 
are assumed. The theoretical prediction is found to exceed the measured value, however, the $\gamma$-ray 
flux is dependent on the viewing geometry.  For example, in the the outer gap model the $\gamma$-ray 
emission decreases with decreasing viewing angle as measured from the rotation axis.
 
Given the constraints on the inclination of the orbital plane of PSR J1023+0038, $\sim 34^{\circ}$ 
to $\sim 53^{\circ}$, it is likely that the line of sight  will pass through edge of the $\gamma$-ray 
beam from the outer gap accelerating region, and that the observed $\gamma$-ray flux will be much 
smaller than $F_{\gamma}\sim 10^{-9}~\mathrm{erg~cm^{-2}~s^{-1}}$. Because the $\gamma$-rays 
associated with edge of the beam will be emitted with an accelerating electric field smaller than 
the typical value described by equation~(\ref{electric}), the energy of the $\gamma$-rays 
will be  smaller  than that given by equation~(\ref{energy}).  
Hence, the recent measurement of $\gamma$-ray radiation in the direction of PSR J1023+0038 can be 
reconciled with the proposed emission model provided that the line of sight passes through the edge 
of the $\gamma$-ray beam. With the typical outer gap geometry, the $\gamma$-ray pulse profile is 
expected to be either a single peak or double peak with a small phase-separation for this viewing 
orientation (c.f. Hirotani 2008).

The secondary pairs produced outside the gap region by the pair-creation process will emit  
non-thermal X-rays via the synchrotron radiation process with a typical energy of $E_{n,X}\sim 2 
(\Gamma/10^{3})^2(B\sin\theta/10^5)~\mathrm{keV}$, where $\Gamma$ is the typical Lorentz factor 
of the secondary pairs, $B$ is the magnetic field near the light cylinder and $\theta$ is the pitch 
angle.  The optical depth of the pair-creation process of a $\gamma$-ray can be estimated as 
$\tau_{X\gamma} \sim n_{s,X}\sigma_{X\gamma}R_{lc}\sim 10^{-2}$ with $n_{s,X}\sim L_{s,x}/(4\pi 
R_{lc}^2cE_{s,X})$ corresponding to the X-ray photon density, where we adopted the observed X-ray 
luminosity $L_{s,X}\sim 10^{31}~\mathrm{erg~s^{-1}}$ and $E_{s,X}\sim 30$~eV (Homer et al. 2006), 
In addition, $\sigma_{X\gamma}\sim \sigma_T/3$ is the cross section of the photon-photon pair creation.
The luminosity of these non-thermal X-rays is estimated as 
\begin{equation}
L_{n,X}\sim \tau_{X\gamma}L_{\gamma}\sim 6\times 10^{32} ~\mathrm{erg/s}. 
\end{equation}
which is consistent with the observations $L_{n,X}\sim 5\times 10^{31} ~\mathrm{erg/s}$ 
(c.f. Homer et al. 2006) 
assuming that we may be observing the edge of the high-energy emission region.  The spectral 
shape below  $E_{n,X}\sim 2$~keV will be expressed by the photon index $\Gamma_{n,X}\sim 1.5$, which 
is near the observed value $\Gamma_{n,X}\sim 1.3$ reported by Homer et al. (2006).

To conclude, it has been shown that the irradiation by $\gamma$-rays from the outer gap can 
facilitate dissolution of an accretion disk within the assumption that the rotation powered 
pulsar is active during a sufficiently low luminosity level in the quiescent state of a soft X-ray 
transient system and the heating leads to the escape of disk material from the system.  The model 
has been applied to the recently born MSP PSR J1023+0038, and we have estimated that the $\gamma$-ray 
luminosity is $\sim 6\times 10^{34}~\mathrm{erg~s^{-1}}$, which is consistent with the irradiated 
luminosity required to explain the heating of the companion star.

Future $\gamma$-ray observations of the pulsed component are needed to confirm that J1023+3008 is 
a $\gamma$-ray emitting pulsar and to place constraints on our theoretical model of the $\gamma$-ray 
emission, its site of production, and its effectiveness in facilitating the removal of the accretion 
disk in this system.

\acknowledgements This work was supported in part by the Theoretical Institute for Advanced 
Research in Astrophysics (TIARA) operated under the Academia Sinica Institute of Astronomy \& 
Astrophysics in Taipei, Taiwan and by NSF AST-0703950 to Northwestern University.

\newpage

\newpage

\newpage
\end{document}